\documentclass[12pt,a4paper]{article}
\usepackage[T2A]{fontenc}
\usepackage[utf8]{inputenc}
\usepackage[english]{babel}
\usepackage{amssymb}
\usepackage{amsmath}
\usepackage{graphicx}
\newcommand{\msun}{\,$M_{\odot}$}
\newcommand{\ergs}{\,erg\,s$^{-1}$}
\newcommand{\kms}{\,km\,s$^{-1}$}

\newcommand{\gcmq}{\,g\,cm$^{-3}$}

\begin{document}

\begin{center}
	\large{\bf Annihilation of positrons from $^{22}$Na in novae}
	
	\vskip 5mm
	\small{N. N. Chugai$^{1}$ and A. D Kudryashov$^{2}$}
	\vskip 5mm
	{\it $^{1}$Institute of Astronomy, Russian Academy of Sciences,
		Pyatnitskaya ul. 48, Moscow, 119017 Russia} \quad  \small{e-mail: nchugai@inasan.ru}\\
	{\it $^{2}$Russian Institute of scientific and technical Information, 
    Russian Academy of Sciences, Moscow}\\
	
\end{center}
 
 \vskip 5mm
\centerline{\bf Abstract} 
\vskip 3mm

We explore for the first time effects of the magnetic field on the escape of $^{22}$Na 
  positrons and on the flux evolution of annihilation 511\,keV line in novae.
It is shown that for the white dwarf magnetic field of $\sim 10^6$\,G the 
 field of the 
  expanding nova shell is able to significantly impede positrons escape and increase the 
  time of the nova emission  in 511\,keV up to hundreds days.
\vskip 3mm 
Keywords: stars --- novae --- nucleosynthesis

\newpage

\section{Introduction}

Nova phenomenon is caused by a thermonuclear runaway (TNR) in a hydrogen shell 
 accumulated on a white dwarf (WD) by an accretion in a binary system with a normal companion 
 (usually red dwarf).
Despite a general understanding of the nova outburst, widely accepted models 
  for major processes involved in the nova phenomenon are lacking.
Particularly, a mixing mechanism of the WD matter with the hydrogen shell that is
 crucial for the TNR is still debatable.
Moreover, it is not clear, whether the nova ejects less or more material than 
 accumulated between consecutive outbursts (Epelstain et al. 2007; Starrfield et al. 2020).
Uncertainties remain also in the description of the shell ejection and shell morphology. 
 
Given the theoretical status, a significant role in the theoretical progress 
  belongs to observations. 
This is supported by two fundamental facts laid in the basis of the nova theory: 
  (i) novae are binary system containing the WD (Kraft 1964);
  (ii) nova shell material is enriched by a factor of 100 with CNO elements 
  (Mustel \& Baranova 1966), which indicates a crucial role of mixing between accreting 
  and WD matter for the nova energetics (Sparks et al. 1976). 
It should be emphasised that apart from novae related to the TNR on  the CO WD
  there is an abundant class of novae related to the TNR on ONe dwarfs (Starrfield et al. 1986). 
This conclusion is prompted by the detection of Ne overabundance in spectra of some novae 
  (Ferland \& Shields 1978).
 Noteworthy, according to the interesting conjecture (Shara \& Prialnik 1994)
  neon novae could arise from CO WD 
 with outer layers significantly enriched by Ne and Mg due to the preceded  
  hydrogen accretion with the rate of $\sim 10^{-6}$\msun\,yr$^{-1}$.
   
The nova TNR results in the synthesis of radioactive isotopes which decay with half life 
  from hundreds of seconds ($^{13}$N) till several years ($^{22}$Na).
As far back as half century Clayton \& Hoyle (1974) emphasised the importance of  
  observations of gamma-lines from novae. 
The isotope composition of nova shell depends on WD type (CO or ONe) and on 
  some poorly fixed parameters, particularly, the WD mass, the accretion rate and 
  the admixing of the WD material into the hydrogen shell.
A detection of gamma-lines from novae would permit us to determine the mass of a certain 
  isotope and thus to obtain useful constraints on outburst models.
  
Highly prospective for gamma-line observations is the isotope $^{22}$Na 
  with the half life of 2.6 years that provides a possibility to apply a long 
  integration time.
The outburst theory predicts a high $^{22}$Na abundance in neon novae, $\sim10^{-3}$ by mass 
  (Denissenkov et al. 2014; Kudryashov 2019), which is by a three order higher compared 
  to that of CO novae.
The $^{22}$Na decay is accompanied by the emission of the 1275\,keV gamma-line, the 511\,keV  
  annihilation line, and three-photon annihilation continuum.
Five neon novae have been observed with the gamma-ray telescope COMPTEL resulting in 
  the upper limit of 
   the $^{22}$Na mass between $3\cdot10^{-8}$\msun\ and $2\cdot10^{-6}$\msun\ based on 
   the $2\sigma$ upper limit of the flux in the 1275\,keV line (Iyudin et al. 1995).
These upper limits do not contradict to theoretical $^{22}$Na abundance in current models 
   of neon novae provided a total ejecta mass is $<3\cdot10^{-5}$\msun.
Subsequently, Iyudin (2010) reported on the COMPTEL detection of 1275\,keV line at the 
  confidence level of 4$\sigma$ from the slow  
   nova V723 Cas of 1975, which is rather unexpected result for these category of novae.

The model luminosity of novae in the annihilation 511\,keV line of $^{22}$Na has been addressed  
  by Gomez et al. (1998) who demonstrated that already a week after the TNR the nova shell 
  becomes transparent for positrons that escape without annihilation.
If this is the case, the possibility to detect 511\,keV line from $^{22}$Na in novae 
  becomes doubtful.
On the other hand, the conclusion about the high transparency of the nova shell for $^{22}$Na 
 positrons is made neglecting a possible magnetic field in the shell.
It may well be that a moderate value of the magnetic field would be enough to essentially impede 
 the positrons escape and thus to prolong the time of a significant nova luminosity in 511\,keV line.
In this case the detection possibility for 511\,keV line from neon novae would become feasible.
  
This paper for the first time explores the issue, to which extent the 
  inclusion of the magnetic field could affect the escape of $^{22}$Na positrons from 
  the shell and the evolution of 511\,keV line flux from the neon nova.
The shell model with the frozen-in magnetic field and the prescription for the annihilation  
   are considered in Section 2. 
The computations of the 511\,keV flux evolution for different values of relevant parameters 
  are presented in Section~3.

\section{Model overview}
\label{sec:mod}

\subsection{Nova shell with frozen-in magnetic field}
\label{sec:shell}

The issue of the shell structure and kinematics for different subclasses of novae 
  currently is not completely clear.
The most natural is the assumption that the ejecta forms in the regime of the optically 
  thick wind (Ruggles \& Bath 1979; Kato \& Hachisu 2007).
The idea is based on the fact that the major stage of the TNR is longer compared to  
  the hydrodynamic time at the WD surface.
Following this arguments we assume that the ejected shell is produced by the wind 
  with the 
  constant mass loss rate and constant velocity of $v \sim 1500$\kms\ that forms at the 
  outer boundary of the expanded envelope with the radius 
  $r_1 \sim 2GM/v^2 \sim 10^{10}$\,cm, where $M \sim 1$\msun\ is the WD mass. 
The convenient wind parameter is the wind kinetic luminosity $L_w = 0.5wv^3$, 
  where $w = \dot{M}/v = 4\pi r^2\rho$ is the wind density parameter. 
The $L_w$ value should be of the order of the Eddington luminosity,   
   $1.26\cdot10^{38}(M/M_{\odot})$\ergs. 
For the standard model we adopt $L_w = 10^{38}$\ergs.
It takes thus $t_w = M_s/\dot{M} \sim 26$ days for the wind to form ejecta with the mass 
 of $M_s \sim 10^{-5}$\msun.
 
For the adopted wind velocity the ejecta perturbation from the 
  binary components are insignificant. 
Indeed, for the typical orbital period of 3.5\,h, the WD mass of $M_1 = 1$\msun, and 
 the red dwarf mass   
   of  $M_2 = 0.5$\msun\ one gets the large semi-axis $a = {9\cdot10^{10}}$\,cm, and 
   orbital velocities of components $v_1 = 150$\kms\ and $v_2 = 300$\kms.
The maximal variation of the wind velocity caused by the WD motion is about 10\% 
  close to the orbital plane.
The variation of the density and velocity caused by the gravitation from the secondary is 
  also small due to high wind velocity.
Below we adopt that the shell expands spherically with the constant velocity.  
Apropos, the neon nova V382 Vel (Takeda \& Diaz 2019) shows the spherical shell    
  expanding with the velocity of 1200\kms.
A possible deviation from the spherical symmetry (cf. Chomiuk et al 2020) is discussued in the 
  final section.

In our picture the hydrogen shell before the outburst resides in the magnetic field 
  of the WD with the average strength of $B_0 = 10^6$\,G.
This value is rather moderate compared to the field of polars and intermediate polars 
  of  $3\cdot10^6 - 2\cdot10^8$\,G. (Ferrario et al. 2020).
The assumption on the magnetic field value is also consistent with the magnetic momentum of 
 the WD of classical nova 1934 DQ Her, viz.,  $(1-3)\cdot10^{32}$\,G\,cm$^3$ 
  (Patterson 1994).
For the WD radius of $\sim4\cdot10^{8}$\,cm this value corresponds to the 
 magnetic field on the WD surface of $(1-3)\times10^6$\,G.
  
The TNR is accompanied by a vigorous convection all over the hydrogen shell
  (Casanova et al. 2018; Starrfield et al. 2020).
The adopted magnetic field ($\sim 10^6$\,G) is weak in the energetic and dynamic  
  respect, so the convection unavoidably results in the creation of the entangled field 
  in the convective shell with the thickness of $h \approx 250$\,km in the case of  
  a ONe WD of 1.25\msun\ (Casanova et al. 2018).
We neglect both the amplification of the magnetic field by the convection and   
  the field dissipation at the small turbulent scales.
The average magnetic field in the convective shell is assumed to be equal to the average 
  field on the WD surface $B_0 = 10^6$\,G.

Our scenario suggests that the nova shell forms in three stages: (i) the TNR accompanied by 
  the convection, (ii) the hydrogen shell expansion, and (iii) the wind outflow.
The envelope expansion at the stage (ii) from the WD radius of $r_0 = 4\cdot10^8$\,cm 
 up to the radius of $r_1 = 10^{10}$\,cm results in the weakening of the 
 frozen-in magnetic field down to $B_1 = B_0(r_0h/r_1^2)$ that is adopted as the initial 
 field of the wind at the level of its formation.
The subsequent evolution of the field in the wind proceeds differently for 
   the radial ($B_r$) and tangential ($B_t$) components.
The radial field decreases as $B_r = B_1(r_1/r)^2$, whereas the tangential as  
  $B_t = B_1r_1^2/(r\Delta r)$, where $\Delta r = vt_w$ is the shell thickness that 
  remains constant.
At the late time $t \gg \mbox{max}(r_1/v, \Delta r/v)$ the entangled field $B$ in 
  the shell is 
  dominated by the tangential components, so we adopt $B = B_t$.
We neglect the shell radial spread-out due to thermal velocities and ignore a possible 
  shell compression by the fast rarefied wind at the nebular stage of the 
  nova evolution ($t \gtrsim 100$\,d).

\subsection{Positron annihilation and escape}
\label{sec:annih}

 The $^{22}$Na isotope decays via the positron emission with the probability of 0.9 and 
   the $e^-$ capture with the probability of 0.1.
 In both cases the daughter nucleus $^{22}$Ne ends up in the excited state 
   with a subsquent emission of 1275\,keV gamma-quantum (Firestone et al. 1999).
Positron energies are distributed in the range of 0 - 546\,keV. 
Our description of positron annihilation takes into account the numerical result that 
 the probability of a free annihilation is $p_f = 0.1$ (Krannell et al. 1976; 
   Leising and Clayton 1987), whereas 90\% of positrons annihilate via the positronium 
   formation at thermal energies.
   
A positron slows down via ionization losses that depend on the matter composition and 
  the ionization degree.
We take into account only hydrogen (X = 0.7) and helium and adopt that the 
 hydrogen ionization fraction is 0.5, while helium is neutral.  
In this case the fraction of free electron per barion is $y_f = 0.35$ and the fraction of 
  bound electrons is $y_b = 0.5$.
The positron slowdown is calculated according to the Bethe formula for ionization losses that is 
  splitted into terms related to neutrals and free electrons (Ahlen 1980)
\begin{equation}
\frac{dE}{d\mu} = -\frac{4\pi e^4N_A}{mc^2\beta^2}\left(y_b\ln{\frac{2mc^2\beta^2\gamma^2}{I}} +
  y_f\ln{\frac{2mc^2\beta^2\gamma^2}{\hbar \omega_0}}\right)\,,
\end{equation}
where $\mu$ is the surface density along the positron path (g\,cm$^{-2}$), $e$ is the elementary charge, 
 $N_A$ is the Avogadro number, $m$ is the electron mass, $c$ is the speed of light,
 $\beta = u/c$ ($u$ is the positron velocity), $\gamma$ is the positron Lorentz factor,
 $I = 29$\,eV is the average ionization potential for the adopted mixture of  
  hydrogen and helium and the ionization fraction, $\omega_0$ is the plasma frequency. 
 
The positron range for a given energy is calculated taking into account the energy distribution 
 (Xiao et al. 2018).
As an illustration, for the initial energy of 200\,keV and 400\,keV the positron range at 
$t = 100$\,d ($\omega_0$ depends on the shell age) is 
 $R = 0.012$\,g\,cm$^{-2}$ and  $R = 0.035$\,g\,cm$^{-2}$, respectively. 
These values differ from the average range of 0.1\,g\,cm$^{-2}$ estimated earlier 
 (Leising \& Clayton 1987). 
The difference presumably is related to the adopted ionization degree. 
Indeed, we find for the neutral gas the comparable value $R \sim 0.1$\,g\,cm$^{-2}$. 
The slowdown due to free electrons is efficient because of the significant energy loss 
in Coulomb collisions.

The escape probability is determined by the ratio of the total surface density along the 
 random walk $\mu$ (g\,cm$^{-2}$) in the shell and the range: 
  $p_{esc} = (1 - p_f)\exp{(-\mu/R)}$.
The $\mu$ value is assumed to be a product of the total geometrical length of the 
  random walk $l$ and the density at the radius of the positron creation. 
An alternative version, based on the average density of the shell, produces the similar result.    
With the probability of ($1-p_{esc}$) the positron radiatively recombines with 
 the branching ratio of 1/4 and 3/4 into the singlet and triplet states, respectively, 
 in line with statistical weights. 
The subsequent kinetics depends on the balance between annihilation rates 
  $t(^1S) = 1.2\cdot10^{-10}$\,s,  $t(^3S) = 1.38\cdot10^{-7}$\,s (Karshenboim 2004), 
  on the one hand, collisional transition rates $^1S~ \rightleftarrows~ ^3S$ and 
  the photoionization rate, on the other hand; these processes are taken into account.
The photoionization might affect the ratio between two- and three-photon annihilation 
  at the early epoch.
However, already the day after the TNR, the photoionization rate $P$ in the wind turns out  
 already small enough, $Pt(^3S) < 0.1$, so at the subsequent stage the ratio between the
 two-photon and three-photon annihilation does not deviate from 1/3.

A frozen-in entangled magnetic field impedes the positrons escape and 
  this should change the evolution of the annihilation line from the $^{22}$Na decay 
  compared to the case without magnetic field.
The positron diffusion in the entangled field is unlike the case of a uniform field. 
In the latter case the diffusion across the magnetic field is strongly inhibited in the 
  absence  of collisions and field perturbations, whereas along the field the positrons can 
  propagate freely. 
In the entagled field the diffusion accros the field can occur because at some coherent 
  length $l_{\mbox{\tiny B}}$ the field significantly changes that results in the positron 
  drift accross the field. 
The diffusion in the entagled field looks therefore like an excursion along the field 
  on the scale $l_{\mbox{\tiny B}}$ and a drift accross the field.
Although there are approaches to the description of the diffusion in the entangled field 
  based on the scale $l_{\mbox{\tiny B}}$ and Larmor radius (Narayan \& Medvedev 2001; 
  Chandran \& Maron 2004) we do not see a possibility to apply these approaches without 
  three-dimensional modelling of the magnetic field and the positron random walk.
   
Here we use the diffusion treatment in terms of the average Bohm diffusion coefficient   
 $D_{\mbox{\tiny B}} = (1/3)R_{\mbox{\tiny L}}u$, where $R_{\mbox{\tiny L}}$ is the 
  Larmor radius and $u$ is the positron velocity.
For the diffusion coefficient we adopt parametric representation   
  $D = \xi D_{\mbox{\tiny B}}$ with the coefficient $\xi \geq 1$.
This approach with $\xi$ in the range of 1...\,10 is adopted, e.g., for the 
   diffusion in the chaotic field when addressing the particle acceleration in supernova 
   remnants (Marcowith et al. 2006).

The characteristic diffusion time for the positron in the expanding shell with 
  the width $d$ is $t_{dif} \sim d^2/D$ while the total surface density 
  ignoring a possible annihilation is $\mu = \rho ut_{dif}$.
This value combined with the positron range $R$ due to ionization losses  
  determines the avearge escape probability of positrons $p_{esc}$ introduced above.
 In the adopted model of the nova shell and the magnetic field evolution for the fixed 
   initial value at the WD surface the probability $p_{esc}$ is determined by the 
   parameter $\xi$ only.
The adopted form of the diffusion coefficient is equivalent to a random walk with the 
 step $l$ being proportional to the shell radius $R_s$.
Indeed, in our model $l \propto R_{\mbox{\tiny L}} \propto B^{-1} \propto R_s$.
It is interesting that in this case the diffusion coefficient formally corresponds 
 to the approximation $\lambda = l/R_s = const$ used earlier in the description of 
 the positrons diffusion in the entangled field of SN~Ia ejecta 
  (Churazov \& Khabibulin 2018).

\section{Results}

A standard model suggests the shell with the mass $M_s = 10^{-5}$\msun, created 
  by the wind with the velocity $v = 1500$\kms\ and the kinetic luminosity $L_w = 10^{38}$\ergs. 
The mass fraction of $^{22}$Na in the shell is $10^{-3}$ and this isotope is assumed to be 
  mixed in the shell homogeneously with the constant fraction in any point.
The wind presumably forms at the radius $r_1 = 10^{10}$\,cm; the result is weakly sensitive 
 to the $r_1$ value.
The average magnetic field at the WD surface is $B_0 = 10^6$\,G. 

The flux of directly escaped 511\,keV quanta from the $^{22}$Na decay in nova at the distance of 
  1\,kpc is shown for $\xi = 1$, 10, and 100, as well as for the case without magnetic field 
  (Fig. 1).
The plot also shows the flux in the 1275\,keV line of $^{22}$Na.
As expected, the magnetic field essentially changes the evolution of the annihilation line, so 
  its detection on the time scale of one year becomes feasible.
Remarkably, in the absence of the magnetic field the flux in the 511\,keV line does not 
  reach the maximal possible value because of the early positrons escape.
For models with the magnetic field and $\xi = 1$, 10, and 100 the fraction of escaped 
  positrons is 0.41, 0.64, and 0.8, respectively.

To remind, we do not take into account the magnetic field amplification by the convection that 
  can be significant. 
Indeed, the MHD 3D-modelling of the solar convection demonstrates that a weak magnetic field 
 can be amplified to the value corresponding the Alfven velocity $V_{\tiny A} \sim 0.45u$ 
 (Cataneo et al. 2003). 
This relation suggests that the convection during the nova outburst with the convective  
  velocity  $u \sim 100$\kms\ (Casanova et al. 2018) in the layer with the density of 
 $\sim3\cdot10^2$\gcmq\ (Denissenkov et al. 2014) is able to amplify the field 
 upto $\sim 10^8$\,G.
For $B_0 > 10^6$\,G positrons trapping will be even more effective.

The effect of other parameters --- the shell mass, the shell velocity, and the width of the   
 convective shell --- in the evolution of the annihilation line from $^{22}$Na decay is 
  shown in Figure 2.
As a fiducial model we take the standard model with $\xi = 10$.
The twice as large shell mass increases the flux and the time of the flux maximum.
The lower expansion velocity by a factor of 1.5 increases the positron trapping and also 
 decreases the time of the flux maximum.
The increase of the width of the convective shell results in the larger field 
  at the late time and therefore more efficient trapping. 
We do not demonstrate the case with the kinetic wind luminosity of $2\cdot10^{38}$\ergs\ 
  because the only effect of this variation is the decrease of the maximum age.
Versions with the different magnetic field are not shown as well because 
  effect of smaller/larger value of the initial field  is similar to the effect of 
  larger/smaller value of the $\xi$ parameter by the same factor.
This suggests that the effect of the field and the diffusion coefficient is actually 
  determined by the combined parameter $B_0/\xi$.

\section{Discussion and conclusions}

The paper has been aimed at the study of effects of the magnetic field in the nova shell on the 
  escape of positrons from the $^{22}$Na decay and on the flux evolution in the annihilation 
  line of 511\,keV. 
We present the model for the positron annihilation in a nova shell that takes into account 
  a magnetic field.
The modelling shows that for the sensible value of the average magnetic field on the 
  WD surface of $10^6$\,G, the frozen-in magnetic field is able to essentially impede 
 the positrons escape and thus to prolong the luminosity of the annihilation line from $^{22}$Na 
  up to hundreds days.
  
 Results, however, depend not only on the strength of magnetic field, but on the 
  diffusion coefficient as well.
 The dimensionless diffusion coefficient $\xi$ used 
   above is in the range $1 \leq \xi \leq 100$, that includes the range of 
     $1 \leq \xi \leq 10$ in models of the diffusive shock particle acceleration 
     constrained by the non-thermal X-ray brightness 
    at the limb of young supernova remnants (Marcowith et al. 2006). 
    
Yet there remains an open question: to which extent an assumption of the homogeneous 
  chaotic magnetic field is applicable to nova shells.
It may well be that the turbulent convection results in the intermittent chaotic field 
  with regions of strong and weak field. 
If this the case and the filling factor of regions with the weak field is significant, 
  the diffusion and escape of positrons could be more efficient compared to the homogeneous 
  case.
Partially this situation is taken into account by the large value of $\xi =100$.
However, actually this does not resolve the issue of a possible intermittency of the 
  chaotic field and the problem requires separate study.
It is noteworthy that a simultaneous detection of 1275\,keV and 511\,keV lines will become 
  a valuable tool for the diagnostics not only of the synthesised  $^{22}$Na mass 
  but also of the parameter $B_0/\xi$ that determines the positron escape.
 
 Shells of historical novae have clumpy structure, e.g., nova 1970 FH Ser 
  (Gill \& O'Brien 2000), and this, at first glance, is in conflict with our model 
  of the spherical shell.
 However, the effect of the clumpy structure on the $^{22}$Na positron escape depends 
   on the stage at which this clumpiness does emerge.
 There is no reason for the clumpiness to appear at the stage of the optically thick wind.
 More naturally to admit that the shell experinces a fragmentation significantly later 
   under the dynamical effect of a high velocity rarefied wind (Cassatella et al. 2004).
 The high velocity wind is formed by the remnant of the hot envelope of the WD due 
   to the radiative acceleration.
 A signature of the wind with the velocity of  $\sim 3000$\kms, versus the 
   main shell velocity of $\sim 1800$\kms, has been observed in nova V1974 Cyg 
   (Cassatella et al. 2004).
 Based on the WR-star wind one expects that the kinetic luminosity of the 
   high-velocity wind ($L_{w,f}$) should be $\sim10^{-2}$ 
   of the radiative luminosity of the WD (Sander et al. 2020), i.e., 
   $L_{w,f} \sim 10^{36}$\ergs\ in our case.
 One can estimate the fragmentation time assuming the fast wind velocity of 
   3000\kms. 
 The major fragmentation mechanism is the Rayleigh-Taylor instability initiated by the 
   shell acceleration driven by the fast wind.
 The fragmentation time scale exceeds the initial phase of the acceleration determined 
  by the shell compression due to the shock propagation.
 For the shell mass of  $M_s = 10^{-5}$\msun\ and the thickness $d = M_s/w$ formed 
   by the wind with the kinetic luminosity $L_w = 10^{38}$\ergs\ the compression 
   by the fast wind takes the time $t_c \sim d/D$, where $D$ is the shock wave speed 
    determined from the momentum conservation 
    $D = (\rho_f/\rho)^{1/2}(v_f - v) = 53$\kms, $\rho_f$ is the density of the fast 
    wind, $v_f$ is its velocity, while the rest of values are related to the nova shell.
 In the considered case $t_c \sim 2$ years.
 This time refers to the initial stage of the nova shell fragmentation that requires 
  somewhat longer time, of the order of $t_f \sim 4t_c$ (Klein et al. 1994), i.e., about 8 years.
 The found fragmentation time indicates that the approximation of the spherical 
  shell on the time scale of one year generally does not contradict to the 
  observed clumpy shells of historical novae.
 
Although novae have been considered as a possible source of Galactic positrons, 
  in effect their contribution to the Galactic 511\,keV gamma-line is unlikely significant 
  even for 100\%  positron escape. 
With the Galactic nova rate of 50 yr$^{-1}$, neon novae fraction of 1/4, nova shell mass of 
   $10^{-5}$\msun, and $^{22}$Na mass fraction of $10^{-3}$ the Galactic production rate of 
   positrons by novae is only of $2\cdot10^{41}$\,s$^{-1}$ that is by a two order lower 
   compared to 
   the Galactic annihilation rate of $5\cdot10^{43}$\,s$^{-1}$ (Siegert et al. 2016).
  
It remains unclear, whether WD of novae are always magnetic with the field   
   $\gtrsim10^6$\,G?
The recent study of the cataclismic variables from the Gaia DR2 within 150\,pc 
  (Pala et al. 2020) shows that 36\% of them posess megagauss fields.
Given the fact that the field of $\sim10^6$\,G is a border value for the cataclismic 
  variables to be classified as an intermediate polar with the accretion on the poles, 
  a significant fraction of WD in cataclismic variables could have magnetic fields 
  $ < 10^6$\,G and thus not show signatures of the intermediate polar.
Yet in this case the magnetic field can be nevertheless 
 strong enough to impede $^{22}$Na positrons escape.
For example, for the field of $10^5$\,G and $\xi = 10$ the situation with the 
   trapping of $^{22}$Na positrons in effect is equivalent 
  to the case of $B_0 = 10^6$\,G  É $\xi = 100$ (Fig. 2) that demonstrates 
   the efficient positron annihilation on the time scale of $\sim 100$ days.
 
 \vspace{1cm}
One of us (N.C.) is grateful to Eugene Churazov and Alexander Getling for stimulating 
 discussions.

\vspace{1.5cm}

%\begin{thebibliography}
%=========================================== 
\centerline{\bf References}
\vspace{0.5cm}

\noindent
Ahlen S.P., Reviews of Modern Physics {\bf 52}, 121 (1980)\\
Casanova J., Jos\'{e} J., Shore S., Astron. Astrophys. {\bf 619}, 121 (2018)\\
Cassatella A., Lamers H. J. G. L. M., Rossi C., et al., Astron. Astrophys. {\bf 420}, 571 (2004) \\ 
Cattaneo F., Emonet T., Weiss N., Astrophys. J. {\bf 588}, 1183 (2003) \\
Chandran B.D.G., Maron J.L., Astrophys. J. {\bf 602}, 170 (2004)\\
Chomiuk L., Metzger B.D., Shen K.J.), arXiv201108751 (2020)\\
Churazov E., Khabibulin I., Mon. Not. R. Astron. Soc. {\bf 480}, 1394 (2018) \\
Clayton D.D., Hoyle F., Astrophys. J. {\bf 187}, L101 (1974)\\
Crannell C.J., Joyce G., Ramaty R., Astrophys. J. {\bf 210}, 582 (1976)\\
Denissenkov P.A., Truran J. W., Pignatari M., et al., Mon. Not. R. Astron. Soc. {\bf442},   
  2058 (2014)\\
Epelstain N., Yaron O., Kovetz A., Prialnik D., Mon. Not. R. Astron. Soc. {\bf 374}, 1449 (2007)\\
Ferland G.J., Shields G.A., Astrophys. J. {\bf 226}, 172 (1978)\\
Ferrario L., Wickramasinghe D., Kawka A., Advances Space Res. {\bf 66}, 1025 (2020)\\
Firestone R.B., Shirley V.S., Baglin C.M. et al.,
  {\it Table of Isotopes} (John Wiley and Sons, New York, 1999)\\
Gill C.D., O'Brien T.J., Mon. Not. R. Astron. Soc. {\bf 314}, 175 (2000) \\
Gomez-Gomar J., Hernanz M., Jose J., Isern J., Mon. Not. R. Astron. Soc. {\bf 296}, 913 (1998)\\
Iyudin A.F., Astron. Rep. {\bf 54}, 611 (2010) \\  
Iyudin A.F., Bennett K., Bloemen H., et al., Astron. Astrophy. {\bf 300},
   422 (1995)\\
Karshenboim S.G., International Journal of Modern Physics A {\bf 19}, 
  3879 (2004)\\
Kato M., Hachisu I., Astrophys. J. {\bf 657}, 1004 (2007)\\
 Klein R.I., McKee C.F., Colella P., Astrophys. J. {\bf 420}, 213 (1994)\\
Kraft R.P., Astrophys. J. {\bf 139}, 457 (1964)\\
Kudryashov A.D., INASAN Science Reports {\bf 3}, 205 (2019)\\
Leising M.D., Clayton D.D.,  Astrophys. J. {\bf 323}, 159 (1987)\\
Marcowith A., Lemoine M., Pelletier G.), Astron. Astrophys. {\bf 453}, 193 (2006)\\
Mustel E.R, Baranova L.I., Soviet Astron. {\bf 10}, 388 (1966) \\
Narayan R., Medvedev M.V., Astrophys. J. {\bf 562}, L129 (2001)\\
Pala A.F., G\"{a}nsicke B.T., Breedt E., et al.,  Mon. Not. R. Astron. Soc. {\bf494}, 
  3799 (2020)\\
Patterson J., Publ. Astron. Soc. Pacific {\bf 106}, 209 (1994)\\
Ruggles C.L.N., Bath G.T.,  Astron. Astrophys. {\bf 80}, 97 (1979) \\
Sander A.A.C., Vink J.S., Hamann W.-R., Mon. Not. R. Astron. Soc. 
    {\bf491}, 2206 (2020)\\
Shara M.M., Prialnik D., Astron. J. {\bf 107}, 1542 (1994)\\
Siegert T., Diehl R., Vincent A.C., et al., Astron. Astrophys. {\bf 595}, 25 (2016)\\
Sparks W.M., Starrfield S., Truran J.W., Astrophys. J. {\bf 208}, 819 (1976)\\
Starrfield S., Bose M., Iliadis C. et al., Astrophys. J. {\bf 895}, 70 (2020)\\
Starrfield S., Sparks W.M., Truran J.W., Astrophys. J. {\bf 303}, 186 (1986)\\
Takeda L., Diaz M., Publ. Astron. Soc. Pacific {\bf 131}, 54205 (2019)\\
Xiao H., Hajdas W., Wu B., et al., Astroparticle Physics, {\bf 103}, 74 (2018)\\

%=====================================================

%\end{thebibliography}

\newpage
\begin{figure}
\centerline{\includegraphics{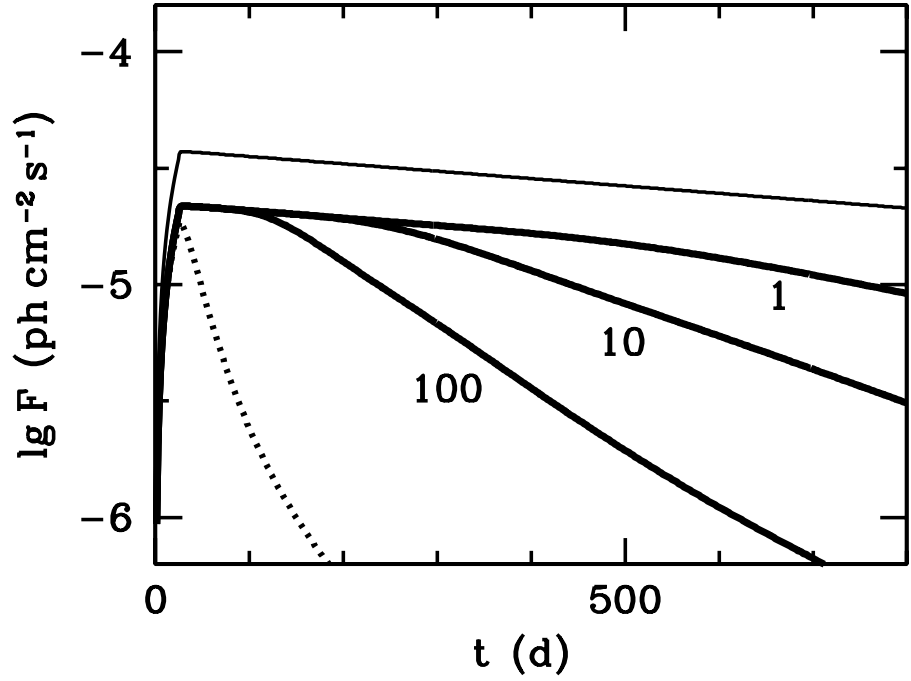}}
\caption{
  Flux of gamma-quanta in 1275\,keV line ({\em thin line}) and in the 
   annihilation line 511\,keV from the $^{22}$Na decay in ONe nova shell at the 
   distance of 1\,kpc. 
   Numbers next to lines indicate the value of $\xi$. {\em Dotted line} shows 
   the model without magnetic field.	
}
\label{fig1}
\end{figure}
\clearpage

\newpage
\begin{figure}
\centerline{\includegraphics{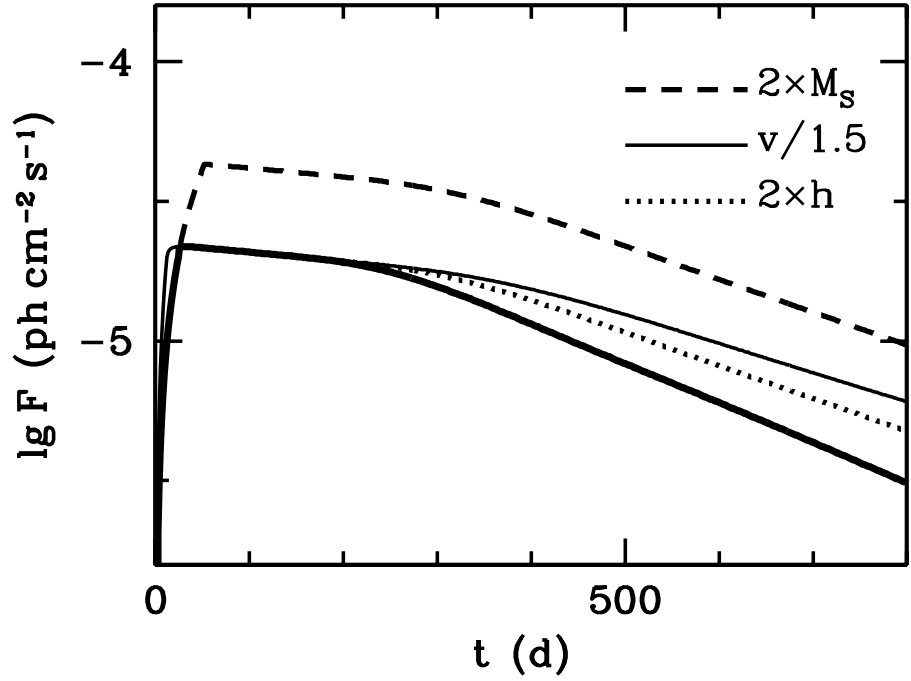}}
\caption{
 Flux in the annihilation line 511\,keV from the $^{22}$Na decay in the 
   standard model with $\xi = 10$ ({\em thick line}) and in models with the variation of 
    a certain parameter (shell mass, velocity, and thickness of the convective shell 
    on the WD surface).
}
\label{fig2}
\end{figure}

%========================================================

\end{document}